\documentclass[sigconf]{acmart}
\usepackage[utf8]{inputenc}
\usepackage[T1]{fontenc}
\usepackage{algorithmic}
\usepackage{graphicx}

\usepackage{textcomp}
\usepackage{xcolor}
\usepackage{makecell}
\usepackage{multirow}
\usepackage{algorithm}
\usepackage{tikz}
\usetikzlibrary{calc, shapes, backgrounds, fit}
\usepackage{varwidth}
\usepackage{calc}
\usepackage{fontawesome}
\AtBeginDocument{%
  \providecommand\BibTeX{{%
    \normalfont B\kern-0.5em{\scshape i\kern-0.25em b}\kern-0.8em\TeX}}}

\newcounter{protostepper}
\newcommand{\protostep}[1]{%
	\def\theprotostepper{#1}%
	\theprotostepper\refstepcounter{protostepper}%
}

\definecolor{gray1}{RGB}{220,220,220}
\definecolor{gray2}{RGB}{200,200,200}

\definecolor{color1}{RGB}{141,211,199}
\definecolor{color2}{RGB}{255,255,179}
\definecolor{color3}{RGB}{190,186,218}
\definecolor{color4}{RGB}{251,128,114}
\definecolor{color5}{RGB}{128,177,211}
\definecolor{color6}{RGB}{253,180,98}
\definecolor{color7}{RGB}{179,222,105}
\definecolor{color8}{RGB}{252,205,229}

\definecolor{colorDark1}{RGB}{228,26,28}
\definecolor{colorDark2}{RGB}{55,126,184}
\definecolor{colorDark3}{RGB}{77,175,74}
\definecolor{colorDark4}{RGB}{152,78,163}
\definecolor{colorDark5}{RGB}{255,127,0}

\tikzset{
	none/.style = {fill=none, rectangle, rounded corners},
	inner/.style = {fill=blue!20, circle, draw=black},
	leaf/.style = {fill=yellow!20, rectangle, rounded corners, minimum width=0.75cm, minimum height=0.75cm, draw=black},
	single/.style = {fill=gray!20, opacity=0.3, rectangle, rounded corners, minimum width=1.0cm, minimum height=0.9cm, draw=black},
	subtree/.style = {fill=green!20, opacity=0.3, rectangle, rounded corners, minimum width=2.0cm, minimum height=1.9cm, draw=black},
	level/.style = {fill=orange!20, opacity=0.3, rectangle, rounded corners, minimum width=5.0cm, minimum height=0.9cm, draw=black},
	quorum/.style = {fill=gray!20, opacity=0.3, rectangle, rounded corners, minimum width=4.0cm, minimum height=1.2cm, draw=black},
	innerMirror/.style = {inner, fill=blue!10, draw=black},
	leafMirror/.style = {leaf, fill=yellow!10, draw=black},
	mixUser/.style = {fill=gray1, rectangle, rounded corners, minimum width=0.75cm, minimum height=0.75cm, draw=black},
	mixMix/.style = {fill=gray2, rectangle, rounded corners, minimum width=0.75cm, minimum height=0.75cm, draw=black},
	base/.style = {rectangle, rounded corners, draw=black,
		minimum width=2.5cm, minimum height=1cm,
		text centered, font=\sffamily},
		empty/.style = {draw=none},
	storage/.style = {base, fill=gray1},
		pics/vhsplit/.style n args = {5}{
		code = {
			\node[text width=0.5cm] (A) at (#1) {};
			\node[anchor=south west,text width=4.0cm] (B) at (A.east) {#3};
			\node[anchor=north west,text width=4.0cm] (C) at (A.east) {#4};
			\node[inner sep=0pt,draw,rounded corners,fit=(A)(B)(C),blend mode=overlay,overlay,fill=#5] (outer) {}; 
			\node[xshift=-2.10cm] (tt) at (outer.center) {#2};  
			\draw (B.north west) -- (C.south west)
			(B.south west) -- (C.north east);    
		}
	},
}

\begin{document}


\title{Concealed Communication in Online Social Networks}

\author{Fabian Schillinger}
\email{schillfa@tf.uni-freiburg.de}
\orcid{0000-0001-8771-8290}
\affiliation{%
  \institution{Computer Networks and Telematics,\\Department of Computer Science,\\ University of Freiburg}
  \streetaddress{Georges-Köhler-Allee 51}
  \city{Freiburg}
  \state{Germany}
  \postcode{79110}
}

\author{Christian Schindelhauer}
\email{schindel@tf.uni-freiburg.de}
\orcid{0000-0002-8320-8581}
\affiliation{%
	\institution{Computer Networks and Telematics,\\ Department of Computer Science,\\ University of Freiburg}
	\streetaddress{Georges-Köhler-Allee 51}
	\city{Freiburg}
	\state{Germany}
	\postcode{79110}
}

%
%
%
%
%


\begin{abstract}
Online social networks are used frequently by many people: Staying in contact with friends and sharing experiences with them is very important. However, users are increasingly concerned that their data will end up in the hands of strangers or that personal data may even be misused. Secure OSNs can help. These often use different types of encryption to keep the communication between the participants incomprehensible to outsiders. However, participants in such social networks cannot be sure that their data is secure. Various approaches show that even harmless-looking metadata, such as the number of contacts of users, can be evaluated to draw conclusions about the users and their communication. These attack methods are analyzed, and existing secure OSNs are examined, whether these attack methods can be utilized to violate the user's privacy. To prevent these privacy attacks, protocols for a secure centralized OSN are developed. Metadata is obscured in the presented OSM and end-to-end encryption is used for secure communication between clients. Additionally, communication channels are concealed using mix networks such that adversaries cannot determine which user is accessing which data or which user is communicating with whom even with access to the server.
\end{abstract}




\maketitle

\newcommand{\concealedAddress}{\textit{concealed address}}
\newcommand{\concealedAddresses}{\textit{concealed addresses}}
\newcommand{\addressKeys}{\textit{address keys}}
\newcommand{\addressKey}{\textit{address key}}

\newcommand{\readAddressKey}{\textit{read address key}}
\newcommand{\writeAddressKey}{\textit{write address key}}
\newcommand{\writeAddressKeys}{\textit{write address keys}}
\newcommand{\ownerAddressKey}{\textit{owner address key}}

\newcommand{\contentKey}{\textit{content key}}

\newcommand{\createAddressMessage}{\textsc{CreateAddress}-Message}
\newcommand{\errorMessage}{\textsc{Error}-Message}
\newcommand{\addressMessage}{\textsc{AddressCreated}-Message}

\newcommand{\createAddress}[1]{\textsc{CreateAddress(#1)}}
\newcommand{\updateAddress}[2]{\textsc{UpdateAddress(#1,#2)}}
\newcommand{\readAddress}[1]{\textsc{ReadAddress(#1)}}
\newcommand{\writeAddress}[2]{\textsc{WriteAddress(#1,#2)}}
\newcommand{\addressCreated}[1]{\textsc{AddressCreated(#1)}}

\newcommand{\mix}{mix network}

\newcommand{\attDegree}{\textit{NS1}}
\newcommand{\attNeighbors}{\textit{NS2}}
\newcommand{\attCount}{\textit{DS1}}
\newcommand{\attType}{\textit{DS2}}
\newcommand{\attSize}{\textit{DS3}}
\newcommand{\attHist}{\textit{DS4}}
\newcommand{\attTimeSeries}{\textit{T1}}
\newcommand{\attTimingCreating}{\textit{T2}}
\newcommand{\attKeyDistAndReencryption}{\textit{T3}}
\newcommand{\attIP}{\textit{CI1}}
\newcommand{\attGeo}{\textit{CI2}}
\newcommand{\attLogs}{\textit{CI3}}
\newcommand{\attControlMessages}{\textit{CI4}}

\newcommand{\attNetwork}{\textit{NS}}
\newcommand{\attData}{\textit{DS}}
\newcommand{\attTiming}{\textit{T}}
\newcommand{\attControl}{\textit{CI}}

\let\proof\sigproof
\let\endproof\sigendproof

\newtheoremstyle{sig}
{}
{}
{\itshape}
{}
{\scshape}
{.}
{.5em}
{#1 #2\thmnote{\quad(#3)}}

\theoremstyle{sig}

\newtheorem{dfn}{Definition}
\newtheorem{hypothesis}{Hypothesis}
\newtheorem{remark}{Remark}

\section{Introduction}
\label{Introduction}
Many online social networks (OSNs) vie for the users' favor. They offer different unique selling propositions to make the user experience as good as possible so that users spend a lot of time on social networks. In many OSNs, money is earned through advertising, the more users are on the network and the longer they stay there, the greater the advertising income. These can be increased even further if personalized advertising is displayed. This is tailored to the respective user and increases the likelihood that they will react to the advertising and thus increases the profit of the network operators. In doing so, users also lose their anonymity. This personalized advertising is only possible by evaluating the profiles, contacts, reactions to messages in the social networks, etc. On the one hand, these practices are a thorn in the eye of many users and they are therefore increasingly avoiding these social networks. Furthermore, it is often not only the operator of a social network that analyzes the personal data, but also third parties, provided they have access to it. These third parties can do harmless things with the data, but they can also display malicious behavior and misuse the collected data for personal theft, for example. and they report this causes a rethinking of many users and they therefore register in social networks that supposedly protect their privacy. This often results in a type of encrypted communication. But apart from the communication, the privacy can be attacked and the sovereignty of the users over their data undermined if the communication is encrypted and user profiles are protected, but the metadata is analyzed.
\subsection*{Our Contribution}
\label{Our Contribution}
Our contribution is a summary of possible privacy leakages in secure and privacy preserving online chats and online social networks. Additionally, an analysis of different privacy preserving OSNs is given, according to the privacy leakages. Following these findings, protocols are presented to create concealed channels between participants in an OSN, which relys on a client-server architecture. These concealed channels work comparable to \mix~. These channels provide end-to-end encrypted communication between two or more participants and further do not leak any evaluable metadata to the service provider of the OSN or another attacker. Using concealed channels different possibilities are presented to provide the functionalities, an OSN should provide. These are, for example, profiles of the users, private messages between two or more participants, and discussion groups. The presented approach of a secure and privacy preserving OSN, then, is analyzed in detail, whether the possible privacy leakages are prevented. 
\subsection*{Organization of the Paper}
The paper is structured as follows: Section~\ref{subsect:privacyAnalysis} summarizes different approaches to analyze metadata in online social networks and encrypted communication channels. Based on these works, possible leackages are displayed in Section~\ref{subsect:privacyLeakages}. Following, in Section~\ref{subsect:onlineSocialNetworksOverview}, different protocols for privacy preserving online chats and online social networks are analyzed, whether they are secure against the leakages. In Section~\ref{sect:proposedProtocolos}, protocols are presented to achieve secure communication. These protocols are used to construct a privacy preserving OSN with encrypted communication. For the presented OSN, then, the privacy preserving functionalities are analyzed and compared to the previously found leakages in Section~\ref{sect:privacyAnalysis}. Finally, Section~\ref{Conclusions} concludes the work and gives an outlook on future work.
\section{Related Work}
\label{sect:relatedWork}
\subsection{Privacy Analysis in Online Social Networks}
\label{subsect:privacyAnalysis}
The attributes users post in their profiles in a OSN can be used to predict the attributes of other users, according to~\cite{Mislove2010You}. The dataset of the Rice University network and the New Orleans Facebook dataset were used. One of the findings was, that friends are likely to share common attributes and these common attributes form groups. Using these findings, on the one hand, it is possible to predict social circles of users. Using the social circles it is, on the other hand, possible to predict attributes of users.\\
In \cite{Schifanella2010Folks}~two datasets of the Online Social Networks Flickr and Last.fm, and in \cite{Aiello2012Friendship}~three different datasets of Flickr, Last.fm, and aNobii were analyzed to predict social links. In these OSNs users can communicate with each other, form groups (of similar topics), create links to other users, and use tags to annotate contents, such as shared pictures. Various observations were made: users that have more contacts tend to be more active regarding tagging and membership of groups, assortative mixing of nodes in the OSNs was detected, and different patterns of topic similarity between neighbors were found. For assortative mixing, i.e. the observation, that nodes tend to be linked to nodes with similar properties, different properties were investigated, for example the degree of nodes, especially the nearest neighbors degree, average number of tags of nearest neighbors, or average amount of groups. Using these observations similarity between users was calculated to predict social links with high accuracy.\\
Privacy leakage through metadata of decentraliced OSNs was examined in \cite{6197506}. Different adversaries with distinct possibilities to access the data were considered. In a centralized setting the service provider could combine all of them. From the metadata of stored content different observations are possible. The size of an object is an indicator for the type, as text is smaller than images or videos. From the structure of a group of elements conclusions can be drawn, e.g. amount of images in a shared album. The modification history of an object can reveal information, e.g. about user status updates or intensity of activity. Other informations can be obtained from access control mechanisms, like encryption headers. Here, header sizes can allow estimating the number of encryption keys. Adding encryption keys or revoking them, can lead to re-encryption of contents and can allow to draw conclusions about changes in the relations between users. From re-using the same key for different objects one can learn about overlapping access rights. The communication flow can lead to more information. From tracking IP addresses conclusions about online times or working habits can be drawn, using geo-ip mapping services routes, locations, and travelling information can be tracked. Access logs from shared content can be used to determine user-groups, ownership, and access patterns. Timing information may be obtained from newly created objects and (re-)distribution of keys. Different control-operations, like login, adding friends, or searches can be observed.\\
Metadata of Twitter was analyzed in \cite{Perez2018You}. A post in Twitter contains 144 fields of metadata which allow drawing conclusions about the owner of such a post, without considering the actual content of the post. Using machine learning approaches owners of posts can be identified from a group of 10.000 users with 96.7\% accuracy. \\
Identifying social circles from network structure and user profile information was the task in~\cite{Mcauley2014Discovering}. Using machine learning approaches, a model was created that accurately identified social circles in Facebook, Google+, and Twitter.\\
Patterns in user behavior can be recognized even with encrypted traffic. In \cite{7265055}~encrypted traffic of different sequences of actions were collected for popular Android apps. A sample sequence for the Facebook app was to tap on the button to write a post, fill the textbox with some random text, and post this message. Analyzing the network flow as set of time series it was possible to predict different patterns even with TLS/SSL encrypted traffic.
\subsection{Privacy Leakages}
\label{subsect:privacyLeakages}
From the findings in Section~\ref{subsect:privacyAnalysis} the following list summarizes possible problems when metadata can be accessed in an Online Social Network where communication is end-to-end encrypted. Single problems may not neccessarily lead to privacy leakages, but the combinations of different problems can lead to severe violations of privacy.
\begin{itemize}
	\item Structure of Network (\attNetwork)
	\begin{itemize}
		\item Degree of Node (\attDegree) - How many contacts does a user have?
		\item Neighbors of Node (\attNeighbors) - How many contacts do the contacts of a user have?
	\end{itemize}
	\item Structure of Data (\attData)
	\begin{itemize}
		\item Count of Objects, Groups, Keys, \textellipsis~(\attCount) - How many contacts, posts, \textellipsis~does a user, group, \textellipsis~have?
		\item Common Objects, Groups, Keys, \textellipsis~(\attType) - Which contacts, posts, comments, \textellipsis~are common between users, groups, \textellipsis?
		\item Size of Objects or Groups (\attSize) - What is the type of an object (text, media, \textellipsis), how many users are in a group?
		\item History of Objects or Groups (\attHist) - How often does an object or group change?
	\end{itemize}
	\item Timing (\attTiming)
	\begin{itemize}
		\item Time Series Pattern (\attTimeSeries) - Are objects or keys accessed in a specific order, especially with a specific timing?
		\item Timing for Creating Objects and Distributing Keys (\attTimingCreating) - Which keys are associated with which objects?
		\item Key Distribution and Re-Encryption (\attKeyDistAndReencryption) - Which keys are associated with which objects and which keys are added or removed when objects are changed?
	\end{itemize}
	\item Control Information (\attControl)
	\begin{itemize}
		\item IP Address Logging (\attIP) - Track user behavior when the same IP accesses content, or different IP addresses access the same content, especially when the time of day is the same across different days.
		\item Geo-IP Mapping (\attGeo) - Where is a user, which locations are associated with a user?
		\item Access Logs for Objects, Groups, or Ownership (\attLogs) - Which user accesses which objects, groups, \textellipsis?
		\item Control Messages and Queries for Login, Friend Request, or Searches (\attControlMessages) - What is a user doing in the OSN?
	\end{itemize}
\end{itemize}

\subsection{Privacy Preserving Online Social Networks}
\label{subsect:onlineSocialNetworksOverview}
There is a various number of schemes that introduce encryption of messages between two or more participants. Some of the schemes are designed for emails, others are designed especially for online message services. Most of the schemes rely on a client-server structure, but there are peer-to-peer approaches, as well:\\
A scheme for encrypted online chats is \textit{Off-the-record} (OTR)~\cite{OTR1}. The scheme uses new session keys $k_i$ for each message $i$. Each key is negotiated through a Diffie-Hellman key exchange between two participants $A,B$ with keys $x_{Ai}, x_{Bi}$, where keys $x_{zi}, x_{zj}$ for $z \in \{A,B\}, i\neq j$ are independent. This introduces perfect forward secrecy for the communication. Possible leakage vectors could be: \attSize and \attTimeSeries~because an adversary could track the sizes and sending times of messages. When a dedicated server is used for the communication, additionally, \attNetwork, \attIP, and \attGeo~can be exploited by the server.\\
A peer-to-peer system for end-to-end encrypted messages between two participants was \textit{Silent Circle Instant Messaging Protocol} (SCIMP)~\cite{moscaritolo2012silent}. Elliptic Curve Diffie-Hellman was used to agree on a shared key for encrypted messages. As SCIMP was a peer-to-peer approach all leakages utilizing a server where mitigated, still the leakage vectors \attNetwork, \attData~and \attControl~could be exploited by a service provider or another malicious relay.\\
\textit{Private Facebook Chat} (PFC)~\cite{robison2012private} introduced end-to-end encrypted chats inside Facebook. The key distribution works by dedicated servers, that use the Facebook authentication mechanisms. Nearly all leakage vectors seem exploitable, except for \attTimingCreating~and \attKeyDistAndReencryption. No matter, whether the Facebook servers or the PFC servers are considered, because an adversary on either of the servers can access all metadata.\\
Multiple peers can communicate securely using the approach described in~\cite{kikuchi2004secure}. A modified Diffie-Hellman protocol is used to find a common secret for the participants with the help of a server. The leakage vectors \attNetwork, \attData~and \attControl~can possibly be exploited, because a server is used to manage the communication and keys.\\
A comparable approach is discussed in~\cite{yang2008design}. Here, elliptic curve Diffie-Hellman is used for the key agreement. Therefore, the same leakage vectors \attNetwork, \attData~and \attControl~can possibly be exploited.\\
In the \textit{Signal} protocol~\cite{SIGNAL_doubleratchet_1} a shared secret between participants is derived from a key chain. The inputs for this chain are found through a modified Diffie-Hellman key exchange. Messages are encrypted and new keys are used for every message. Still, the leakage vectors \attNetwork, \attData~and \attControl~are possible, when the server is attacked.\\ 
\textit{Threema}~\cite{threema} allows end-to-end encryption of messages. Communication between two peers is encrypted using a shared key. Messages in groups are encrypted for each peer individually, using their public keys. Larger files are encrypted using a symmetric key, which is encrypted with each participant's public key. When the server is attacked, all leakage vectors \attNetwork, \attData, \attTiming, and \attControl could be exploited. \\
Another end-to-end encrypted online chat is presented in~\cite{schillinger2019end}. Messages are encrypted using symmetric AES encryption. The neccessary keys are encrypted for all participants of a chat using their public RSA keys. A chat can contain arbitratily many participants and the amount of participants can be changed, then new AES keys are generated and distributed. There are various possible leakage vectors, when the server is attacked: all from \attNetwork, \attData, \attControl, and \attTiming, because all neccessary data to manage the OSN are stored in plaintext on the server.\\
\textit{Pretty Good Privacy} (PGP)~\cite{rfc4880} and \textit{S/MIME}~\cite{rfc5751} are methods for the encryption of emails. In both, public key encryption is used to encrypt a symmetric key for every recipient of an email. The content of an email is encrypted, using the symmetric key. All following answers use the same keys. The main difference is in the verification of public keys: PGP constructs a trust system between participants, whereas S/MIME uses X.509 certificates. Possible leakage vectors are: \attNetwork~and \attData, because recipients are known in plaintext. Leakages from \attControl~are possible, if the adversary is one of the involved mail servers.\\
Other approaches introduce privacy into Online Social Networks:\\ 
\textit{FlyByNight}~\cite{Lucas:2008:FMP:1456403.1456405} introduces client-side based encryption of content for Facebook. Each user has a password that is used to encrypt a private key for the key database of the system. Messages between participants are encrypted using their public keys. Proxy cryptography is used when more than two participants communicate. As flyByNight is an extension to Facebook all leakage vectors could be used when considering their servers. When considering only the servers of flyByNight, still, \attNetwork, \attTiming, \attControl, \attCount, and \attSize~could be used, because the server manages all keys and messages.
A decentralized OSN is \textit{Safebook}~\cite{5350374}. Social circles from the real life are used to construct trust relationships. Each node is surrounded by those structures, which are called matryoshkas. This is used to provide data storage and communication privacy. Another layer is a peer-to-peer network that enables application services, such as lookup. The internet is the transport layer in the scheme. Because of the peer-to-peer approach, most of the leakages are prevented, or at least very unlikely. E.g. to exploit \attNetwork, \attData, or \attTiming, the trusted peers have to attack the user. Attacks through \attControl~are unlikely because of the peer-to-peer structure.\\
In~\cite{cutillo2009privacy} another decentralized OSN is presented. Again, real life trust relationships are utilized for trustworthy connections within the network. Multihop routing between trusted peers is used as an anonymization technique. Privacy leakages are unlikely, because for \attNetwork, \attData, or \attTiming~trusted peers have to attack the user.\\
In the OSN \textit{Persona}~\cite{baden2009persona} users define who can access their information. Attribute-based encryption is used to share secrets within groups of participants that have at least one attribute in common. Further, each user owns a key-pair, such that the public key can be used to encrypt content specifically for this user. Although communication data is encrypted, various usable metadata may accumulate on the server, therefore, exploits of \attNetwork, \attData, \attTiming, and \attControl~seem possible.\\
\textit{Snake}\cite{barenghi2014snake} is an OSN which is written in HTML5 and JavaScript. It uses the WebCrypto API to encrypt messages between peers. One can not conclude which peers communicate, because addresses are masked inside the database. When users establish a friendship they agree on a shared key, and addresses to send and receive messages. These addresses change with every new message. Therefore, exploiting \attNetwork~and\attData~seems to be not possible, when considering communication between peers, but \attTiming~and \attControl~could be exploited. When a user logs in the server has to prove the neccessary encrypted data and exploits of \attCount, \attSize, and \attHist~can be possible.
\subsection{Mix Networks}
\label{subsect:mix}
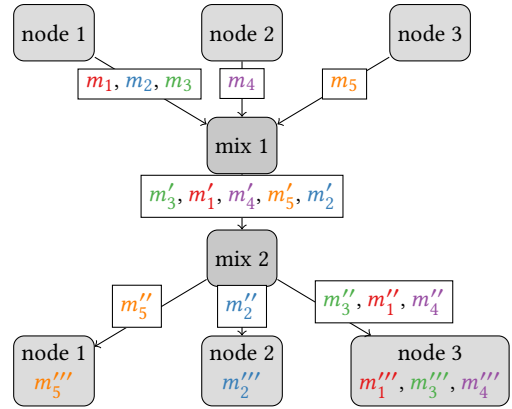
\begin{figure}[htb]
	\centering	
	\begin{tikzpicture}[every node/.style={fill=white, draw=black}, align=center]
		
		\node (node1) [mixUser] at (0, 0) {node 1};
		\node (node2) [mixUser] at (2.5, 0) {node 2};
		\node (node3) [mixUser] at (5, 0) {node 3};
		
		\node (mix1) [mixMix] at (2.5,-1.5) {mix 1};
		\node (mix2) [mixMix] at (2.5,-3.0) {mix 2};
		
		\node (node4) [mixUser] at (0,-4.5) {node 1\\\textcolor{colorDark5}{$m'''_5$}};
		\node (node5) [mixUser] at (2.5,-4.5) {node 2\\\textcolor{colorDark2}{$m'''_2$}};
		\node (node6) [mixUser] at (5,-4.5) {node 3\\\textcolor{colorDark1}{$m'''_1$}, \textcolor{colorDark3}{$m'''_3$}, \textcolor{colorDark4}{$m'''_4$}};
		
		\draw[->] (node1) -- (mix1) node[pos=0.4] {\textcolor{colorDark1}{$m_1$}, \textcolor{colorDark2}{$m_2$}, \textcolor{colorDark3}{$m_3$}};	
		\draw[->] (node2) -- (mix1) node[pos=0.4] {\textcolor{colorDark4}{$m_4$}};		
		\draw[->] (node3) -- (mix1) node[pos=0.4] {\textcolor{colorDark5}{$m_5$}};
		\draw[->] (mix1) -- (mix2) node[pos=0.4] {\textcolor{colorDark3}{$m'_3$},
			\textcolor{colorDark1}{$m'_1$},
			\textcolor{colorDark4}{$m'_4$},
			\textcolor{colorDark5}{$m'_5$},
			\textcolor{colorDark2}{$m'_2$}};
		\draw[->] (mix2) -- (node4) node[pos=0.4,left] {\textcolor{colorDark5}{$m''_5$}};
		\draw[->] (mix2) -- (node5) node[pos=0.4] {\textcolor{colorDark2}{$m''_2$}};
		\draw[->] (mix2) -- (node6) node[pos=0.4,right] {\textcolor{colorDark3}{$m''_3$}, \textcolor{colorDark1}{$m''_1$}, \textcolor{colorDark4}{$m''_4$}};

		\begin{scope}[on background layer]
		\end{scope}
	\end{tikzpicture}
	\caption{A \mix. The nodes send encrypted messages to a mix. The mix collects multiple messages, reorders them, decrypts them the first time and determines the next mix, then the messages are forwarded to the next mix. Again, the messages are collected, reordered, decrypted, and forwarded to the receiver, which then, decrypts them the last time.}
	\label{fig:mixnet}
\end{figure}

In \cite{chaum1981untraceable} mix networks are presented. A \mix~is a routing protocol where a message is sent to a proxy, called mix, that forwards the message to another mix or the recipient. When layered encryption is used between the participants the path of a message becomes hard to trace. This allows to achieve anonymity of the sender, of the recipient, or both. Consider a network with sender $S$, recipient $R$, and mixes $M_1, M_2, \dots, M_n$ with public encryption keys $s, r, m_1, m_2, \dots, m_n$ and a message $N$.\\
Anonymity of the sender is achieved by encrypting the message to $n = \{\{\{\{N\}_r\}_{m_n}\dots\}_{m_2}\}_{m_1}$. $n$ is sent to $M_1$, who decrypts the message and forwards it to $M_2$, and so on. Finally, $R$ receives $\{n\}_r$ from $M_n$. Because each recipient of the message only knows the participant before, and after, the flow of the message is concealed and $R$ does not get to know that $S$ was the sender.\\
Anonymity of the recipient is provided, when the recipient generates a return address. The return address is an encrypted sequence of mixes that the sender has to use, where only the first mix is known to the sender. The recipient sends this sequence to the sender, via the \mix.\\
By combining both schemes, both, sender and recipient of a message can remain anonymous to each other.\\
Different methods are needed, that outsiders cannot track messages through a \mix, such as removing of duplicate messages. This would allow an attacker to find a connection between a received message and the next mix, because the duplicate has to be sent to the exact same receiver. Further, messages have to be modified by a mix to prevent comparing incoming with outgoing messages. Additionally, messages at a mix have to be collected and either forwarded at random or together with other messages and the forwarding procedure has to produce a different ordering of messages.
In Figure~\ref{fig:mixnet} a \mix~is displayed, where three nodes send messages via two mixes.
\section{Proposed Protocols}
\label{sect:proposedProtocolos}
The following protocols ensure private communication within a Online Social Network, based on a client-server architecture. No metadata is leaked when users communicate. This is achieved through concealed communication channels.
\subsection{A Concealed Secure Channel}
\label{sect:secureChannel}
In a client-server model a concealed channel between two clients uses the server. Such a channel can be created when one of the clients has a concealed channel to an address on the server, which is known to the other client and accessed through a concealed channel. This address is called \concealedAddress.
\begin{dfn}[Concealed Address]
	A \concealedAddress~is a tuple $(c, p_R, p_W, p_O)$, where $c$ is a unique address at the server. $p_R, p_W, p_O$ are values to prove that one is allowed to read or write messages to this address, or that one is the owner of the address. 
\end{dfn}
A client can create a \concealedAddress~by sending a \createAddressMessage~to the server. The \concealedAddress~provides an address on the server where clients can read messages from or write messages to if the according proofs are provided. Another proof determines ownership of the \concealedAddress. This means there are three corresponding values, called \addressKeys~for a \concealedAddress: $p_R$, $p_W$, and $p_O$.
\begin{dfn}[Address Key]
	An \addressKey~is a value $p_X \in \mathbb{G}$, where $X \in \{R,W,O\}$ and $\mathbb{G}$ is a finite cyclical group of primal order $q$ with a generator $g$, such that $p_X = g^x \mod q$, with $x\in\mathbb{N}$. The \addressKey~can have a wildcard-value $\ast$, when $x=0$.
\end{dfn}
Clients can prove that they are allowed to read messages from the \concealedAddress~by using the $p_R$~value, which is called \readAddressKey. $p_W$, the \writeAddressKey, is used to verify if a client is allowed to write messages to it. It can be proven that the client is the owner using the \ownerAddressKey $p_O$. For the verification whether a client is allowed to read messages the server generates two random values $m_R, r_R$ and computes ${c_0 \equiv g^{r_R} \mod q}$ and ${c_1 \equiv g^{R r_R}\cdot m_R \mod q}$. $c_0, c_1$ are provided to the client. If the client can calculate ${m_R \equiv \frac{c_1}{c_0^{r_R}} \equiv \frac{g^{R r_R}\cdot m_R}{g^{R r_R}} \mod q}$ he is allowed to receive the messages from the server. The procedure, see Algorithm~\ref{algo:proveRead}, is the same for proving the permission to write to the address or proving ownership of the \concealedAddress. The proof, that a client is allowed to read or write messages to a \concealedAddress~is only a verification against the server. The other clients, that write messages to the \concealedAddress~or read messages from it, may use a key for the encryption and decryption of messages, which is unknown to the user. This key is called \contentKey. 
\begin{dfn}[Message Key]
	A \contentKey~is a value $k$, which is used as a symmetric key for the encryption and decryption of messages stored at a \concealedAddress.
\end{dfn} The \contentKey~and \addressKeys~are independent from each other. They can be negotiated between the participants or distributed in person and later changed between messages.
Using multiple \concealedAddresses~concealed communication channels, comparable to a \mix~(see Section~\ref{subsect:mix}), can be established. A participant of the OSN, that wants to provide a mix network generates multiple \concealedAddresses, where the \writeAddressKeys~are $\ast$. This allows everybody knowing the addresses to write messages to it. When the messages are encrypted with, either the public key of the mix or a symmetric key, only known to the mix and the sender, they can be decrypted by the mix, only. These messages, then, contain another encrypted message for another mix or a recipient and an address, where it has to be written to. Messages are collected by the mix, decrypted, reordered, merged if possible, and then written to the specified addresses. Messages can be merged, when they have the same target address. Then, they are concatenated and written as a single message. For increased security, duplicate messages are deleted. Using this construction, a communication between a sender and a receiver can be established, where the sender is anonymous, or the receiver is anonymous, or both are anonymous to each other. An exemplary message flow is depicted in Algorithm~\ref{algo:concealedChannel}. To reduce the amount of messages stored to the server, mixes can delete messages, after they are forwarded. However, to prevent attacks with duplicate messages, the mixes have to wait a certain amount of time, depending on how many messages they receive and have to forward. When a path is used that achieves anonymity of the sender, this anonymous sender can prove ownership of an address to the server. This allows participants of the OSN to exchange addresses. E.g. a user $A$ can create a \concealedAddress~$c_A$ with some \addressKeys. $A$ can send the \ownerAddressKey~via a concealed channel to $B$. Then, $B$ can prove ownership of the \concealedAddress, via another concealed channel. Exchanging \concealedAddresses~between users and mixes cannot be tracked by the server or another user. 
	\begin{algorithm}[htb]
	\centering
	\begin{tabular}{
			@{}p{2em}
			@{}p{4em}
			@{}p{\linewidth-6em}
			@{}}
	\protostep{1}. 
	\label{algo:proveRead:step1}
	& \textbf{Client} & sends \readAddress{$c$}~from \concealedAddress $c$ \\[0ex]
	& \multicolumn{2}{p{\linewidth-1.4em}}{\hspace*{-0.6em}\textbf{\boldmath If $p_R = \ast$:}}\\[0.0ex]
	\protostep{2a}. 
	& & sends messages in \concealedAddress \\[0.0ex]
	& & terminates. \\[0.8ex]
	& \multicolumn{2}{p{\linewidth-1.4em}}{\hspace*{-0.6em}\textbf{\boldmath If $p_R \neq \ast$:}}\\[0.0ex]
	\protostep{2b}. 
	& \textbf{Server} & chooses random $m_R$ and $r_R$  \\[0ex]
	& & calculates $c_0 \equiv g^{r_R} \mod q$\\[0ex]
	& & calculates $c_1 \equiv {p_R}^{r_R} \cdot m_R \mod q$\\[0ex]
	& & sends $(c_0, c_1)$ to the client \\[0.8ex]
	& \multicolumn{2}{p{\linewidth-1.4em}}{\hspace*{-0.6em}\textbf{\boldmath If client knows $R$:}}\\[0.0ex]
	\protostep{3b.a}.
	& \textbf{Client} & calculates $m_R \equiv \frac{c_1}{c_0^{r_R}} \equiv \frac{{g^R}^{r_R} \cdot m_R}{{g^R}^{r_R}}  \mod q$ \\
	& & sends $m_R$ to the server \\[0ex]
	\protostep{4b.a}.
	& \textbf{Server} & sends messages in \concealedAddress \\[0.0ex]	
	& & terminates. \\[0.8ex]
	& \multicolumn{2}{p{\linewidth-1.4em}}{\hspace*{-0.6em}\textbf{\boldmath If client does not know $R$:}}\\[0.4ex]
	\protostep{3b.b}.
	& \textbf{Client} & sends another value \\[0ex]
	\protostep{4b.b}.
	& \textbf{Server} & denies access\\[0ex]	
	& & terminates. \\[0.0ex]	
\end{tabular}\\[0ex]
	\caption{This algorithm is used to verify that a client is allowed to read messages from a \concealedAddress. The value $p_R \equiv g^R \mod q$ is known to the server, $R$ is a private key, that is only known to authorized clients. The verification procedure for writing messages or claiming the ownership over the \concealedAddress works similar. Here the values $p_W \equiv g^W \mod q$ and $p_O \equiv g^O \mod q$ are used, where $W,O$ are private keys. Possible messages for Step~\ref{algo:proveRead:step1}, are shown in Table~\ref{tab:possibleMessages}.}
	\label{algo:proveRead}
\end{algorithm}

	\begin{algorithm}[htb]
	\centering
	\begin{tabular}{
			@{}p{2em}
			@{}p{4em}
			@{}p{\linewidth-6em}
			@{}}
	\protostep{1}. 
	& \textbf{Client} & creates a new symmetric key $k$ for encryption \\[0ex]
	& & sends \createAddress{$k$} \\[0ex]
	\protostep{2}. 
	& \textbf{Server} & creates new address $c$, chooses random $r,w,o$\\[0ex]
	& & calculates $p_R \equiv g^{r} \mod q p_W \equiv g^{w} \mod q, p_O \equiv g^{o} \mod q$ \\[0ex]
	& & encrypts $(c,p_R,p_W,p_O)$ with $k$ to $\{c,p_R,p_W,p_O\}_k$ \\[0ex]
	& & sends $\{c,p_R,p_W,p_O\}_k$ \\[0ex]
	\protostep{3}.
	& \textbf{Client} & decrypts $\{c,p_R,p_W,p_O\}_k$ \\[0ex]
	& & chooses random $R'$ and calculates calculates $p'_R \equiv g^{R'} \mod q$ or sets $p_R = \ast$ \\[0ex]
	& & chooses random $W'$ and calculates calculates $p'_W \equiv g^{W'} \mod q$ or sets $p_W = \ast$ \\[0ex]
	& & chooses random $O'$ and calculates calculates $p'_O \equiv g^{O'} \mod q$ or sets $p_O = \ast$ \\[0ex]
	& & sends \updateAddress{$c$}{$p'_R,p'_W,p'_O$}\\[0ex]
	\protostep{4}.
	& \textbf{Server} & verification procedure according to Algorithm~\ref{algo:proveRead} \\[0ex]	
	& \multicolumn{2}{p{\linewidth-1.4em}}{\hspace*{-0.6em}\textbf{\boldmath If proof is incorrect:}}\\[0.4ex]
	\protostep{5a}.
	& \textbf{Server} & sends \errorMessage, terminates \\[0.8ex]
	& \multicolumn{2}{p{\linewidth-1.4em}}{\hspace*{-0.6em}\textbf{\boldmath If proof is correct:}}\\[0.4ex]
	\protostep{5b}.
	& \textbf{Server} & stores $(c,p'_R,p'_W,p'_O)$, sends \addressCreated{$c$}.\\[0ex]	
	& & terminates \\[0ex]	
\end{tabular}\\[0ex]
	\caption{This algorithm is used to create a new \concealedAddress. The values $r,w,o$, chosen by the server are replaced by values known to the client only.}
	\label{algo:createAddress}
\end{algorithm}

\begin{algorithm}[htb]
	\centering
	\begin{tabular}{
			@{}p{2em}
			@{}p{4em}
			@{}p{\linewidth-6em}
			@{}}
		\protostep{1}. 
		& \textbf{Sender \boldmath $S$} & encrypts $m$ to $m' = \{m\}_r$ \\[0ex]
		& & chooses random $l$ \\[0.0ex]
		& & creates list $\mathit{destinations} = (R)$\\[0.0ex]
		& \multicolumn{2}{p{\linewidth-1.4em}}{\hspace*{-0.6em}\textbf{\boldmath For 1 to $l$ do:}}\\[0.0ex]
		& & sets $\mathit{receiver} = $~last element of $\mathit{destinations}$\\[0.0ex]
		& & sets $\mathit{dest} = $~address of $\mathit{receiver}$\\[0.0ex]
		& & creates message $m = (\mathit{message}: m', \mathit{destination}: \mathit{dest})$ \\[0.0ex]
		& & chooses random $M_i$ \\[0.0ex]
		& & appends $M_i$ to $\mathit{destinations}$\\[0.0ex]
		& & encrypts $m$ to $m' = \{m\}_{m_i}$\\[0.0ex]
		& \multicolumn{2}{p{\linewidth-1.4em}}{\hspace*{-0.6em}\textbf{\boldmath EndFor}}\\[0.0ex]
		& & sends $m'$ to the address of the last element of $\mathit{destinations}$\\[0.0ex]
		\protostep{2}. 
		& \textbf{Mix} & decrypts $m'$ to $(\mathit{message}: m', \mathit{destination}: \mathit{dest})$ \\[0.0ex]
		& & sends $m'$ to $\mathit{dest}$ \\[0.0ex]
		\protostep{3}. 
		& \multicolumn{2}{p{\linewidth-1.4em}}{\hspace*{-0.6em}\textbf{...}}\\[0.0ex]
		\protostep{4}. 		
		& \textbf{Receiver} & decrypts $m'$ to $m$ \\[0.0ex]
	\end{tabular}\\[0ex]
	\caption{The sender $S$ wants to send a message $m$ to the recipient $R$, with address $c_R$ and public key $r$. The addresses $c_{i,1}, c_{i,2}, \dots, c_{i,j}$ of mix $M_i$ with public key $m_i$ and are known to $S$.}
	\label{algo:concealedChannel}
\end{algorithm}

\begin{table}
	\caption{Possible messages a user can send to the server to create, update, read or write \concealedAddresses.}
	\label{tab:possibleMessages}
\begin{tabular}{lp{13.5em}}
	\toprule
	Type & Description\\
	\midrule
	\createAddress{$k$} & Creates a new \concealedAddress, communication between the client and the server are encrypted using the key $k$ \\
	\updateAddress{$c$}{$R,W,O$} & Updates a \concealedAddress~$c$ with $R,W,O$ are parameters for proofs for reading, writing, and ownership \\
	\readAddress{$c$} & Reads messages from the \concealedAddress~$c$\\
	\writeAddress{$c$}{$m$} & Writes a new (encrypted) message $m$ to a \concealedAddress $c$\\
	\bottomrule
\end{tabular}
\end{table}
\subsection{User Information and Postings}
\begin{figure*}[htb]
\centering	
\begin{tikzpicture}[every node/.style={fill=white, draw=black}, align=center]
	\node (profile)	[storage, rectangle split, rectangle split parts=2] at (-6,0) 
			{
				\textbf{(a6ebc5, $\ast$, g\textsuperscript{9050e7}, g\textsuperscript{fe3485})}
				
				\nodepart{second}
				\begin{tabular}{c|lllll}
				0 & Name: & \multicolumn{3}{l}{Sample User} \\ 
				\hline 
				1 & Public key: & \multicolumn{3}{l}{yFpQlYLlx3eiTJ29}  \\ 
				\hline 
				2 & \{Date of Birth: & address: & e211fb\}\textsubscript{1c031a}, & key: & $k_1$ \\ 
				\hline 
				3 & \{Residence:  & address:& 85cd71\}\textsubscript{1c031a}, & key: & $k_1$ \\ 
				\hline 
				4 & \{School: & address:& 4f8e3d\}\textsubscript{b9f1cf}, & key: & $k_2$ \\ 
				\hline 
				5 & \{Contactlist 1: & address:& 54deaf\}\textsubscript{8eaeeb}, & key: & $k_3$  \\ 
				\hline 
				6 & \{Contactlist 2: & address:& 777c9d\}\textsubscript{3312cd}, & key: & $k_4$ \\ 
				\hline 
				7 & \{Postings: & address:& 76703f\}\textsubscript{1c031a}, & key: & $k_1$ \\ 
				\hline 
				8 & \{Keys: & address:& a9a7d6, & key: & 418530\}\textsubscript{6770da}  \\ 
				\hline 
				9 & \multicolumn{4}{l}{\textellipsis} \\ 
				\end{tabular} 
			};

	%
			
				\node (keys)	[storage, rectangle split, rectangle split parts=2] at (-7,-4.5) 
						{
							\textbf{(a9a7d6, g\textsuperscript{418530}, g\textsuperscript{271812}, g\textsuperscript{c18a4c})}
							
							\nodepart{second}
							\begin{tabular}{c|llllll}
							0 & \{keyId: & $k_1$ & address: & 8c5620 & key: & dd11e5\}\textsubscript{418530}\\
											\hline 
							1 & \{keyId: & $k_2$ & address: & 65889b & key: & 57b042\}\textsubscript{418530}\\
											\hline 
							2 & \{keyId: & $k_3$ & address: & b66f9a & key: & def981\}\textsubscript{418530}\\
											\hline 
							3 & \{keyId: & $k_4$ & address: & 300a03 & key: & b451cc\}\textsubscript{418530}\\
											\hline 
							4 & \multicolumn{3}{l}{\textellipsis}
							\end{tabular}
			
						};
					
					\node (keys1Users)	[storage, rectangle split, rectangle split parts=2] at (1,-6) 
					{
						\textbf{(8c5620, g\textsuperscript{dd11e5}, g\textsuperscript{b1a72d}, g\textsuperscript{8eb72e})}
						
						\nodepart{second}
						\begin{tabular}{c|llll}
							0 & \{user: & 7c5d03 & key: & 1c031a\}\textsubscript{4209d1}\\
							\hline 
							1 & \{user: & f0d082 & key: & 1c031a\}\textsubscript{c7b3df}\\
							\hline 
							2 & \{user: & 74344e & key: & 1c031a\}\textsubscript{c35fb0}\\
							\hline 
							3 & \multicolumn{3}{l}{\textellipsis}
						\end{tabular}
					};
						
						\node (postingsInBetween)	[storage, rectangle split, rectangle split parts=2] at (4,-0.5) 
						{
							\textbf{(76703f, g\textsuperscript{1c031a}, g\textsuperscript{9d30eb}, g\textsuperscript{897047})}
							
							\nodepart{second}
							\begin{tabular}{c|llllll}
								0 & \{address: &a59f4a & key: & 161995\}\textsubscript{1c031a}\\
							\end{tabular}
						};
						
						\node (postings)	[storage, rectangle split, rectangle split parts=2] at (4,-3) 
												{
													\textbf{(a59f4a, g\textsuperscript{161995}, g\textsuperscript{ddd5d9}, g\textsuperscript{517f14})}
													
													\nodepart{second}
													\begin{tabular}{c|llll}
													0 &\{content: &"posting content"\}\textsubscript{1c031a}, &key:& $k_1$\\
													\hline 
													1 &\{content: &"posting content"\}\textsubscript{75fb62}, &key:& $k_5$\\
													\hline 
													2 &\{content: &"posting content"\}\textsubscript{1c031a}, &key:& $k_1$\\
													\hline 
													3 &\{content: &"posting content"\}\textsubscript{1c031a}, &key:& $k_1$\\
													\hline 
													4 &\multicolumn{3}{l}{\textellipsis}
													\end{tabular}
												};
												
%

			\node (postingsInTable)	[draw=colorDark3, fill=none, rounded corners=7pt, minimum width=8.6cm, minimum height=0.4cm] at (-5.75,-1.3) { };
			\node (postingsInIntermediateTable)	[draw=colorDark3, fill=none, rounded corners=7pt, minimum width=5.4cm, minimum height=0.4cm] at (4.2,-0.8) { };
			
			\node (keysInTable)	[draw=colorDark4, fill=none, rounded corners=7pt, minimum width=8.6cm, minimum height=0.4cm] at (-5.75,-1.7) { };
			\node (keysInKeysTable)	[draw=colorDark4, fill=none, rounded corners=7pt, minimum width=7.4cm, minimum height=0.4cm] at (-6.8,-3.95) { };
			
			\draw[->, colorDark3] (postingsInTable.east) -- (postingsInBetween.west);
			\draw[->, colorDark3] (postingsInIntermediateTable.south) -- (postings.north);
			
			\draw[->, colorDark4, dashed] (keysInTable.south) -- (keys.north);
			\draw[->, colorDark4] (keysInKeysTable.east) -- (keys1Users.west);

\end{tikzpicture}
\caption{A user profile solely relying on \concealedAddresses. The construction allows a user to hide its personal information, except for the name. The fields are encrypted using keys $k_1,\dots, k_4$, which are stored in a \concealedAddress, which is not directly, i.e. in plaintext, linked from the profile. When a user receives the key $6770da$, it can decrypt the entry containing the keys. Then, the \concealedAddress, with the keys can be accessed. Each key, then, is encrypted, using a user-specific key. Using these keys, the fields from the profile can be decrypted and their \concealedAddresses~can be read.}
\label{fig:userProfile}
\end{figure*}
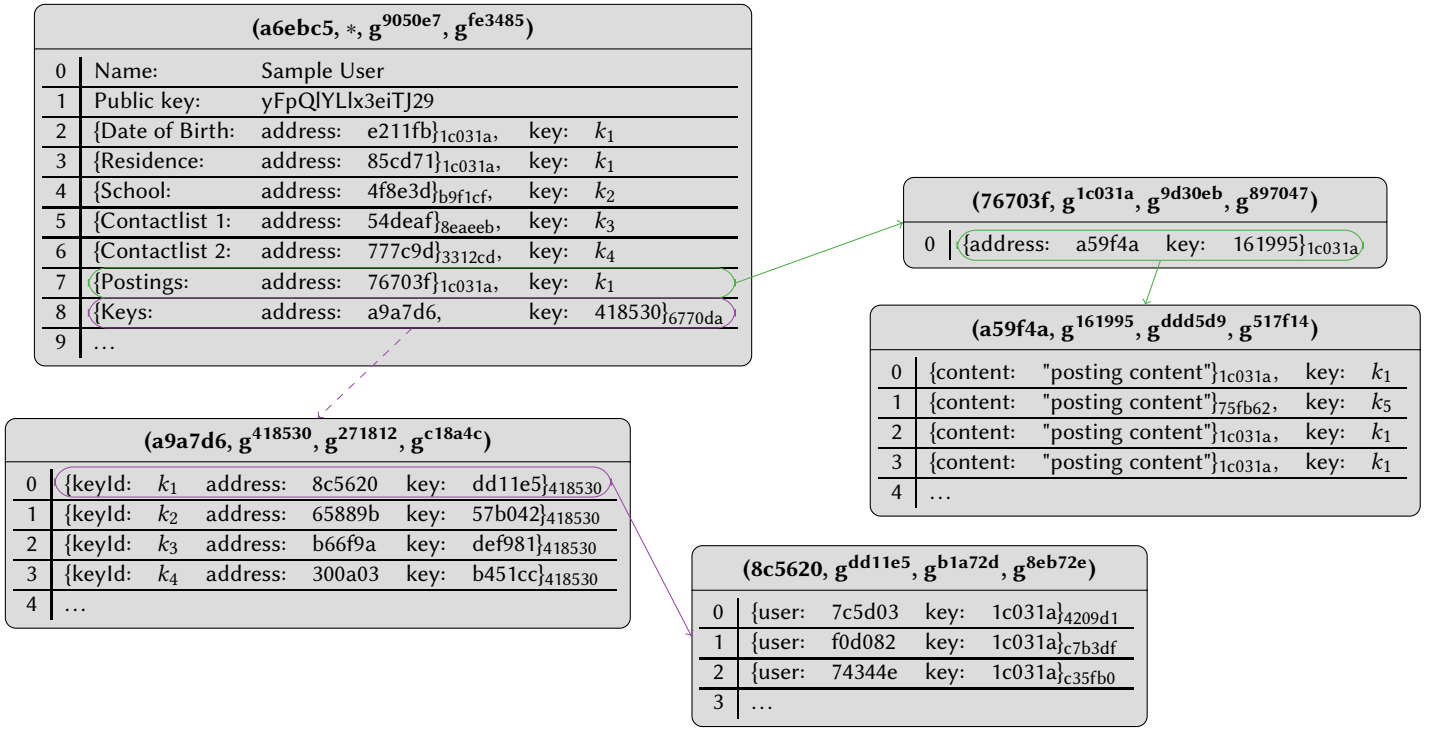

Sharing personal user information, like the date of birth or residence, is part of social networking. In many OSNs users can create profiles to share such user information. In our approach it is possible to create a user profile using \concealedAddresses. The user can decide, how the profile is stored at the server: a user profile can consist of a single \concealedAddress, containing all the information a user wants to share. This information can be encrypted or stored as plaintext at the \concealedAddress. The user then decides who receives the \concealedAddress, the \readAddressKey, and, in case of encrypted information, who receives the \contentKey. Storing the user profile as a whole at the server is possible but not recommended: on the one hand, a user has to update the whole profile, when some information is changed. On the other hand, no fine-grained access rights can be defined. Then, a user cannot share some of the information with everybody public, while other information is shared with specific contacts, only. To achieve this, a user can create a \concealedAddress, where the \readAddressKey~is $\ast$. This \concealedAddress~is used to store all addresses of the profile information, like a \concealedAddress~containing the public keys of the user, so other users can verify signatures or send encrypted messages, another \concealedAddress~for the date of birth, another \concealedAddress~for the residence, and so on. Then each linked \concealedAddress~can have a unique \addressKey. To make the tracking of user information through linked addresses difficult, each linked address itself can contain another linked address, which is encrypted. This prevents an adversary from learning which addresses are connected.\\ A user may want to share postings. These can contain personal experience, for example, pictures from the last holidays. This information can be stored at the user profile, like personal information, or the user can link a single, or multiple \concealedAddresses. Each address can contain a list of posts. This allows the user to create different information feeds, where each feed can have different keys, and therefore different access rights.\\ The construction of a user profile using \concealedAddresses~allows not only to post personal profile information and user posts, but any type of information: the user may share a calender with fine grained access control over each appointment, different blogs or vlogs, picture albums, or wikis. The user can create different \concealedAddresses, where the \writeAddressKey~is $\ast$. These addresses work as different pinboards, where other users can post messages visible to the user and everybody knowing the correct \readAddressKey. Further, the user can use private \concealedAddresses~to store notes, bookmarks, or other private information, like passwords. For every information the user can create a new \concealedAddress, a new \addressKey, and a \contentKey~to be able to define the access rights every time. Using \concealedAddresses, the user is able to create collections of keys. These collections can be shared with single users or with groups of users. An exemplary user profile is displayed in Figure~\ref{fig:userProfile}.
\subsection{Personal Messages and Group Messages}
Personal messages or group messages are possible between two or multiple participants. One user creates a \concealedAddress~and shares the \readAddressKey~and \writeAddressKey~with the other participants. The \contentKey~can be chosen by this user and shared, using the public keys of the other participants. Then, the participants can write messages to this \concealedAddress~and read the incoming messages. Whenever a new participant is added to a chat the \readAddressKey, \writeAddressKey, and \contentKey~is shared with this user, via an encrypted message, using the public key. When a user is removed from the chat the owner of the address, knowing the \ownerAddressKey, has to change the \readAddressKey~and \writeAddressKey, and a new \contentKey~is generated and shared with the remaining participants.
\subsection{Groups}
A group in an OSN is a collection of informations, messages, posts, pictures, etc., comparable to a user profile, with the difference, that everything in the group may be accessed by multiple users. Additionally, some parts of the group my be changed by some participants. A group can be constructed like a user profile. One \concealedAddress~is used as a collection of different linked \concealedAddresses, containing the different information shared in the group. Because each \concealedAddress~can have a unique key, again a fine grained access system can be created, where some information in the group can be accessed by all members of the group, where other information is hidden to some members. At some addresses, members of the group can add or edit information, at other addresses only a specified group of administrators of the group can change information. The necessary keys, can be shared between the authorized users, via personal messages or via shared \concealedAddresses. The construction of groups and user profiles, using \concealedAddresses, allows users to join groups, without exposing their information. Then, they just read the \concealedAddresses~of that specific group. When users want to be visible within the group, they can publish their profile's address or their name, together with a signature, within the group.
\subsection{Finding and Verifying Participants, Exchanging Keys}
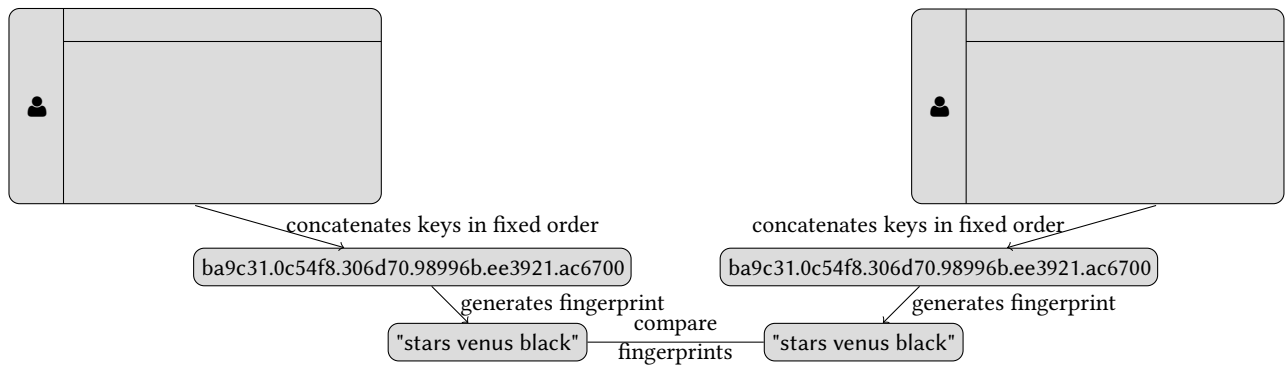
\begin{figure*}[htb]
	\centering	
	\begin{tikzpicture}[align=center]
		\path pic (userA) {vhsplit={0,0}
				{\faUser}
				{\textbf{\boldmath User $A$}}
				{
					\begin{tabular}{ll}
						\multicolumn{2}{l}{Own Public Keys:}\\
						 & \faKey\textsubscript{$A$}:~ba9c31, \faPencil\textsubscript{$A$}:~0c54f8\\
						\multicolumn{2}{l}{Other Public Keys:}\\
						 & \faKey\textsubscript{$B$}:~306d70, \faPencil\textsubscript{$B$}:~98996b\\
						 & \faKey\textsubscript{$C$}:~ee3921, \faPencil\textsubscript{$C$}:~ac6700\\
					\end{tabular}}
				{gray1}};
		\node (userAempty)	[empty] at (2,-2.15) { };
			
		\node (concatenatedA)	[storage, minimum width=0.5cm, minimum height=0.5cm,] at (5,-3) {ba9c31.0c54f8.306d70.98996b.ee3921.ac6700};
		\node (fingerprintA)		[storage, minimum width=0.5cm, minimum height=0.5cm,] at (6,-4) {"stars venus black"};
		
		\draw[->]     (userAempty) -- node[xshift=2.3cm] {concatenates keys in fixed order} (concatenatedA);
		\draw[->]     (concatenatedA) -- node[xshift=1.5cm] {generates fingerprint} (fingerprintA);
		
		\path pic (userB) {vhsplit={12,0}
			{\faUser}
			{\textbf{\boldmath User $B$}}
			{
				\begin{tabular}{ll}
					\multicolumn{2}{l}{Own Public Keys:}\\
					& \faKey\textsubscript{$B$}:~306d70, \faPencil\textsubscript{$B$}:~98996b\\
					\multicolumn{2}{l}{Other Public Keys:}\\
					& \faKey\textsubscript{$A$}:~ba9c31, \faPencil\textsubscript{$A$}:~0c54f8\\
					& \faKey\textsubscript{$C$}:~ee3921, \faPencil\textsubscript{$C$}:~ac6700\\
			\end{tabular}}
			{gray1}};
		\node (userBempty)	[empty] at (15,-2.15) { };
		
		\node (concatenatedB)	[storage, minimum width=0.5cm, minimum height=0.5cm,] at (12,-3) {ba9c31.0c54f8.306d70.98996b.ee3921.ac6700};
		\node (fingerprintB)		[storage, minimum width=0.5cm, minimum height=0.5cm,] at (11,-4) {"stars venus black"};
		
		\draw[->]     (userBempty) -- node[xshift=-2.3cm] {concatenates keys in fixed order} (concatenatedB);
		\draw[->]     (concatenatedB) -- node[xshift=1.5cm] {generates fingerprint} (fingerprintB);
		
				\draw[-]     (fingerprintA) -- node[xshift=0.0cm] {compare\\fingerprints} (fingerprintB);
	\end{tikzpicture}
	\caption{Users $A$ and $B$ verify their keys. Both users concatenate the public keys in a fixed order, like using ascending ids, then fingerprints are generated. Both fingerprints are compared via a secure channel, like meeting in public. If both fingerprints are equal $A$ and $B$ know that their communication is secure.}
	\label{fig:verifyPublicKeys}
\end{figure*}

A user has the freedom in constructing a profile, such that no other user can find him. This can be achieved, when no profile information is published, or all of the profile fields are encrypted by the user. Further, if a user has encrypted profile fields, these cannot be found through the OSN. An attribute-based search for participants can be prevented. Nevertheless, when a user publishes some fields, an attribute-based search is possible. When a user does not want to be found through the OSN the only way to find him is by receiving the needed \concealedAddresses~and keys through a different channel. A channel can be via phone or by meeting in person. The user can generate a fingerprint, like a QR code or a textual representation of the \concealedAddresses~and keys via a method described in~\cite{dechand2016empirical}. \\This method can be used to verify public keys of users, as well: when participants $A$ and $B$ want to verify their keys, they call a procedure that concatenates all the public keys of $A$ and $B$ in a predefined order, like by ascending id of the public keys. Then a fingerprint is generated by both users $A$ and $B$ and compared. When both fingerprints are equal, they can be sure that the public keys are equal and no third party has injected a false key. When $A$ and $B$ trust each other, they can use this procedure to verify keys of other participants, as well, and construct a web of trust this way. This procedure is displayed in Figure~\ref{fig:verifyPublicKeys}. As soon as two participants have established a trusted connection they can exchange encrypted messages through a \concealedAddress. This allows them to further exchange and verify keys. When users have a verified \concealedAddress~to exchange messages they can work as mixes for each other. By forwarding messages to other mixes, or participants of the OSN.\\
When a user created a profile comparable to Figure~\ref{fig:userProfile}, other users can find him through the OSN, because the server can read the Name-entry in the profile address, because the \concealedAddress~can be read by anybody and the entry is not encrypted. A similar public field can be used in groups, or by mixes, as well.
\label{sect:privacyAnalysis}
\section{Conclusions}
\label{sect:Conclusions}
We have presented a summary of different possible privacy leakages within secure online social networks and a discussion whether different approaches for privacy preserving OSNs are possibly vulnerable to these leakages. Either, the OSNs are possibly vulnerable or are using a peer-to-peer architecture. Therefore, we presented a way to construct privacy preserving, encrypted, concealed channels in a client-server architecture. No evaluable metadata is generated when using these channels, according to the previous findings of possible privacy leakages. Using these concealed communication channels different protocols are presented to provide the full functionality of OSNs. These functionalities contain: a user profile, where profile fields, contact lists, pinboards, etc. can be enrypted and hidden for the server, an attacker, or any third party. Further, private messages between two or more participants of the OSN are provided. Another presented functionality are groups. A group can contain message boards, calendars, pinboards, etc. The groups can be accessed by multiple participants. Some of the participants may be only able to read contents, whereas, other participants can produce contents and publish them in a group. Protocols to verify public keys via a secure channel are discussed, as well as the interchange of concealed addresses.\\
One of the main advantages of the OSN, however, can be considered as the main weakpoints as well: profile information or concealed channels can be hidden from the server, to prevent the server or a third party from evaluating this information. On the other hand, this prevents any user from searching these informations through the server. This means that two participants have to meet via a different channel in order to exchange the information they need in order to ultimately be able to conduct a private communication via the OSN. However, once this hurdle has been overcome, any further verification of other OSN participants can also take place via this communication channel. Further research, therefore, may address this point to make the OSN a bit more user-friendly.
\begin{acks}
The authors acknowledge the financial support by the Federal Ministry of Education and Research of Germany in the framework of SoNaTe (project number 16SV7405).
\end{acks}

\bibliographystyle{ACM-Reference-Format}
\bibliography{verschleiertesosn}

\end{document}